\def\ts     {\thinspace}
\def\kms    {\ifmmode{{\rm \ts km\ts s}^{-1}}\else{\ts km\ts s$^{-1}$}\fi}
\def\msol   {\ifmmode{{\rm M}_{\odot}}\else{M$_{\odot}$}\fi}
\def\lsol   {\ifmmode{{\rm L}_{\odot}}\else{L$_{\odot}$}\fi}
\def\zsol   {\ifmmode{{\rm Z}_{\odot}}\else{Z$_{\odot}$}\fi}
\def\etal   {{\rm et\ts al.}}
\def\aco    {\ifmmode{^{12}{\rm CO}(J\!=\!1\! \to \!0)}\else{$^{12}${\rm CO}($J$=1$\to$0)}\fi}
\def\bco    {\ifmmode{^{12}{\rm CO}(J\!=\!2\! \to \!1)}\else{$^{12}${\rm CO}($J$=2$\to$1)}\fi}
\def\cco    {\ifmmode{^{12}{\rm CO}(J\!=\!3\! \to \!2)}\else{$^{12}${\rm CO}($J$=3$\to$2)}\fi}
\def\ci     {\ifmmode{{\rm C}{\rm \small I}}\else{C\ts {\scriptsize I}}\fi}
\def\hi     {\ifmmode{{\rm H}{\rm \small I}}\else{H\ts {\scriptsize I}}\fi}
\def\hh     {\ifmmode{{\rm H}_2}\else{H$_2$}\fi}
\def\cone {\ifmmode{{\rm C}{\rm \small I}(^3\!P_1\!\to^3\!P_0)}
     \else{C\ts {\scriptsize I}{\small$(^3\!P_1\!\to^3\!\!\!P_0)$}}\fi}
\def\ctwo {\ifmmode{{\rm C}{\rm \small I}(^3\!P_2\!\to^3\!P_1)}
     \else{C\ts {\scriptsize I}{\small$(^3\!P_2\!\to^3\!\!\!P_1)$}}\fi}
\def\cij {\ifmmode{{\rm C}{\rm \small I}\,(^3P_i\to^3P_j)}\else{C\ts {\scriptsize I}\,{\small$(^3P_i\to^3P_j)$}}\fi}
\def\cii    {\ifmmode{{\rm C}{\rm \small II}}\else{C\ts {\scriptsize II}}\fi}
\def\tex {\ifmmode{{T}_{\rm ex}}\else{$T_{\rm ex}$}\fi}
\def\tmb {\ifmmode{{T}_{\rm mb}}\else{$T_{\rm mb}$}\fi}
\def\tkin {\ifmmode{{T}_{\rm kin}}\else{$T_{\rm kin}$}\fi}
\def\microns {\ifmmode{\mu{\rm m}}\else{$\mu$m}\fi}
\def\nhh   {\ifmmode{n({\rm H}_2)}\else{$n$(H$_2$)}\fi}
\documentclass[twocolumn]{aa}
\usepackage{graphicx}
\begin{document}
 \title{Gas and Dust in the Cloverleaf Quasar at Redshift 2.5}

   \author{A. Wei\ss
          \inst{1}
          \and
          C. Henkel
          \inst{2}
          \and
          D. Downes
          \inst{3}
          \and
          F. Walter
          \inst{4}
          }

   \offprints{We use $H_{\rm 0} = 75$ \kms\,Mpc$^{-1}$ and $q_{\rm 0} = 0.5$}
   \institute{IRAM, Avenida Divina Pastora 7, 18012 Granada, Spain
         \and
             MPIfR, Auf dem H\"ugel 69, 53121 Bonn, Germany
          \and
             IRAM, 300 rue de la Piscine, 38406 St-Martin-d'H\'eres, France
         \and
             NRAO, PO Box O Socorro, NM, 87801 USA 
             }

   \date{}

   \abstract{
   We observed the upper fine structure line of neutral carbon, \ctwo\,
   ($\nu_{\rm rest} = 809$\,GHz), the \cco\, line ($\nu_{\rm rest} = 345$\,GHz) and the 1.2\,mm continuum
    emission from H1413+117 (Cloverleaf quasar, $z=2.5$) using the IRAM 
   interferometer. Together with the detection of the lower fine structure
   line (Barvainis \etal\, 1997), the Cloverleaf quasar is now
   only the second extragalactic system, besides M\,82, where
   {\it both} carbon lines have convincingly been detected. Our
   analysis shows that the carbon lines are optically thin and have an
   excitation temperature of \tex\,$\approx$\,30\,K. 
   CO is subthermally excited and the observed line luminosity
   ratios are consistent with \nhh\,$\approx$\,10$^{3-4}\,{\rm
   cm}^{-3}$ at \tkin\,=\,30--50\,K. Using three independent methods
   (\ci, dust, CO) we derive a total molecular gas mass (corrected for magnification) of
   $M(\hh)$\,$\approx$\,1.2\,$\pm$\,0.3\,$\times$\,10$^{10}\,\msol$. 
  Our observations suggest that the molecular disk extends beyond the region seen in
CO(7--6) to a zone of more moderately excited molecular gas that
dominates the global emission in \ci\, and the low $J$ CO lines.

   }

   \maketitle
%

\section{Introduction}

Detections of large amounts of dust and gas in distant quasars have 
opened the possibility to study molecular gas properties
in the early epoch of galaxy formation and to provide fundamental
constraints on galaxy evolution. To date, CO emission has been detected
in more than a dozen quasar host galaxies with $z$\,$>$\,2, 
the record holder being at $z$\,=\,6.4 (Walter \etal\,
\cite{walter03}, Bertoldi \etal\, \cite{bertoldi03}). 
Molecular gas masses in excess of $10^{10}$\,\msol\, have led to the
hypothesis that the tremendous far--infrared (FIR) luminosities
($>$\,$10^{12}$\,\lsol) of these objects are not only powered by active
galactic nuclei (AGN) but also by major starbursts which might be
forming cores of elliptical galaxies or bulges of massive spiral galaxies. 

\noindent Due to its strong magnification by gravitational lensing H1413+117 
is a relatively strong emitter in various molecular lines and
therefore a prime target to study the physical conditions of the
molecular gas at high redshift in great detail. It has been detected 
in the CO 3--2 (Barvainis \etal\, \cite{barvainis94}, 
Wilner \etal\, \cite{wilner95}, Alloin \etal\,
\cite{alloin97}, Barvainis \etal\, \cite{barvainis97}, B97
hereafter), 4--3, 5--4 (B97) and 7--6 (Alloin \etal\, \cite{alloin97}, Yun \etal\, \cite{yun97}) 
transitions, the lower fine structure line of atomic
carbon, \cone, (B97) and in HCN (B97, Solomon \etal\,
\cite{solomon03}).

\noindent In this letter we report on the detection of the upper fine structure line
of neutral carbon, \ctwo, which adds an important piece of information: 
unlike CO, cool \ci\, can be described by a 3-level system. This allows us to derive 
its excitation and to constrain the physical gas conditions by
observations of the \cone\, and \ctwo\, transitions only. We compare our
results from \ci\, with estimates based on thermal dust emission and CO.

\section{Observations and Results}

Observations were carried out with the IRAM Plateau de Bure
interferometer during 2 nights in March 2003. The dual frequency
setup was used to observe the \cco\, and  \ctwo\,
transitions towards H1413+117 (RA 14:15:46.23, Dec 11:29:44.0; J2000) 
simultaneously. The receivers were 
tuned in single sideband mode (SSB) at 97.191 GHz and in dual sideband mode
at 224.478 GHz/227.478 GHz. 
We used the standard D configuration (6 antennas) which results in  
synthesized beams of $7\farcs5\times4\farcs4$ and
$3\farcs5\times1\farcs7$ for the 3\,mm and 1\,mm  bands respectively.
At these resolutions the source is unresolved. 
Typical SSB system temperatures were $\approx$\,120\,K and $\approx$\,300\,K at 3\,mm
and 1\,mm. Flux and passband calibration were obtained on
MWC\,349. The nearby sources 1413+135 and 1423+146 were used as secondary flux and 
phase calibrators. We estimate the flux density scale to be accurate to about 10\%.

\begin{figure}
   \centering
   \includegraphics[width=7.9cm]{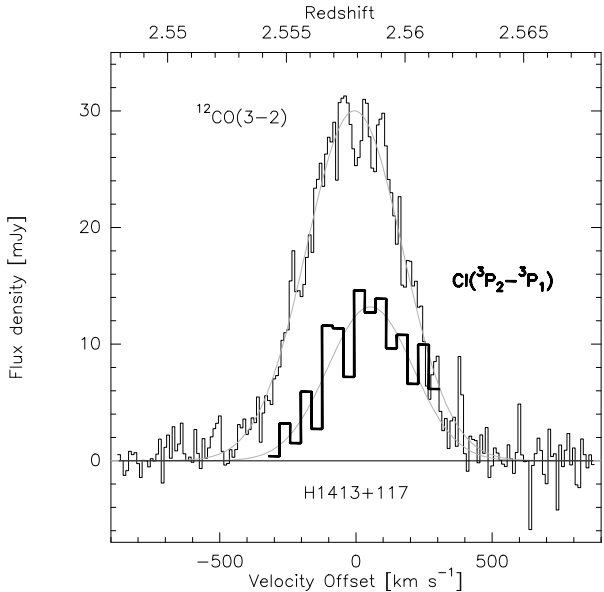}
   \caption{Spectra of the \cco\ and \ctwo\, lines towards H1413+117,
   superposed on their Gaussian fit profiles (see Table~\ref{linepara}
   for parameters). The velocity scale
   is relative to the tuning frequencies of 97.191 and 227.478\,GHz
   corresponding to $z=2.5579$.}
   \label{lines}
   \end{figure}

\begin{figure}
   \centering
   \includegraphics[width=8.5cm]{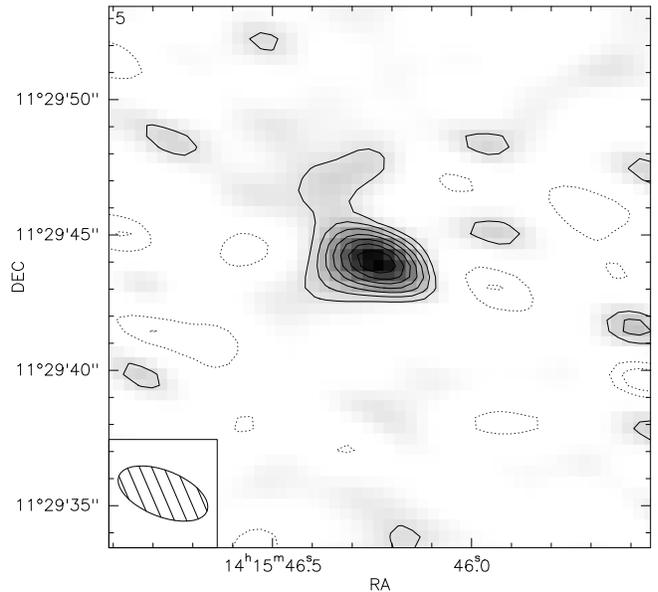}
   \caption{Integrated \ctwo\, intensity distribution in H1413+117. 
Contours~are~plotted~at~$-1.5$,~$-1.0$ and~from~1.0~to 4.0 in steps of
   0.5~Jy\,beam$^{-1}$\,\kms. The rms noise is 0.5~Jy\,beam$^{-1}$\,\kms, the 
   peak flux density is 4.3~Jy\,beam$^{-1}$\,\kms. The continuum has 
  been subtracted.}
   \label{1mmcont}
\end{figure}

\noindent The data were processed to give data cubes with a velocity resolution of
10\,\kms\ (3.24\,MHz) at 3\,mm and 40\,\kms\ (29.7\,MHz) at 1\,mm. 
A 3\,mm continuum image was generated by averaging the line emission--free channels. 
At 1\,mm the continuum image was computed from the
image sideband. From the 1\,mm signal sideband data we generated a
continuum free \ctwo\, spectral line cube by subtracting the 1\,mm continuum
image. The final results are presented in Figs.~\ref{lines}\,\& \ref{1mmcont}.

\noindent The \cco\, transition ($\nu_{\rm rest}$\,=\,345.796\,GHz) is 
detected with high signal to noise (Fig.~\ref{lines}). 
Line parameters for the \cco\, transition are similar to those 
reported by Wilner \etal\,(\cite{wilner95}) and B97. 
The high quality of our \cco\, spectrum allows us to
determine the redshift of the molecular gas 
(as traced by CO) with high accuracy: $z$\,=\,2.55784\,$\pm$\,0.00003.

\noindent The \ctwo\ line ($\nu_{\rm rest}$\,=\,809.342\,GHz, M\"uller \etal\, \cite{mueller01}) is clearly
detected with a peak flux density of $S_\nu$\,=\,13.2\,$\pm$\,2.9\,mJy 
(Figs.~\ref{lines}\,\&\,\ref{1mmcont}). Due to
the limited bandwidth of 512\,MHz (670\,\kms)  
our observations lack the red line--wing. However, this does not affect the determination of
the peak line intensity  since the zero power level is well 
defined in interferometric observations. Table~\ref{linepara}
summarizes the line parameters derived from Gaussian fits.
The \ctwo\, line ($z$\,=\,2.5585\,$\pm$\,0.0001) is displaced by
$\approx$\,+60\,\kms\, relative to \cco\, -- opposite to the displacement of the \cone\, transitions
found by B97 ($\approx$\,--42\,\kms). 
While the difference between the CO and \cone\, redshift might be
attributed to low level baseline instabilities in the 30\,m data and 
the poor signal to noise ratio, the reason for the shift in the \ctwo\, line is 
unclear. Gravitational amplification should not alter the center frequency for \ci\, unless
its distribution is different from CO, which we consider unlikely (see
below). The apparent line shift needs to be confirmed by
higher--sensitivity observations. If confirmed, this shift is most
likely due to an opacity effect.

\noindent For the dust flux, we derive an upper limit for the 3\,mm 
($\lambda_{\rm rest}$\,=\,870\,\microns) continuum of 
$S_{97.2\,{\rm GHz}}$\,$<$\,1.5\,mJy (3\,$\sigma$). For the 1\,mm 
($\lambda_{\rm rest}$\,=\,375\,\microns) continuum
we find $S_{224.5\,{\rm GHz}}$\,=\,7.5\,$\pm$\,0.6\,mJy.
We obtained an additional 250\,GHz flux measurement using the MAMBO
117 channel bolometer array at the IRAM 30\,m telescope with 
$S_{250\,{\rm GHz}}$\,=\,16\,$\pm$\,2\,mJy, consistent with a
previous measurement (Barvainis \etal\, \cite{barvainis95}). 
The radio--IR spectral energy distribution (SED) of H1413+117 is shown in Fig.\,\ref{sed}.
 
   \begin{table*}
      \caption[]{Observed line parameters.
}
         \label{linepara}
            \begin{tabular}{l c c c c c c c}
            \hline
            \noalign{\smallskip}
            Line & $\nu_{\rm obs}$ & $S_\nu$ & $\Delta
      V_{\rm FWHM}$ & $I$& $V^{\mathrm{a}}$ & $L'$/10$^{10}$ & Ref.
            \\
                 & [GHz] & [mJy] & [\kms] & [Jy \kms]&[\kms] & [K
      \kms\, pc$^2$] &\\
            \noalign{\smallskip}
            \hline
            \noalign{\smallskip}
            
 \cco            & 97.1928     &  30.0 $ \pm$ 1.7     & 416 $\pm$  6   &
                   13.2 $\pm$ 0.2 &  $-5.6$ $\pm$ 2.4&18.0$\pm$0.3 &
 this paper\\
 \ctwo           & 227.4376  &  13.2 $ \pm$ 2.9     & 368 $ \pm $ 25 
 & 5.2 $ \pm$ 0.3  & 53 $\pm$ 20 &1.30$\pm$0.07 & this paper\\
\cone             & 138.351  &  7.7 $\pm$ 0.8  & 430 $ \pm $ 46 
 & 3.6 $ \pm$ 0.4  & $-47$ $\pm$ 24 &2.4$\pm$0.3 & B97\\
            \noalign{\smallskip}
            \hline
           \end{tabular}
\begin{list}{}{}
\item[] Quoted errors are statistical errors from Gaussian
            fits. Systematic calibration uncertainty is 10\%  
\item[$^{\mathrm{a}}$] The velocity is given relative to z=2.5579, the
      mean redshift of the CO lines determined by B97.
\item[]CO fluxes for 3--2, 4--3, 5--4, 7--6 are: 13.2, 21.1, 24.0, 47.3
[Jy \kms]; all values except for 3--2 are from B97.
\item[]CO line luminosities $L'$ are: 18.0, 16.1, 11.7, 10.9 [10$^{10}$\,\lsol]
\end{list}
\end{table*}
\section{Analysis}
\subsection{Atomic Carbon}
\subsubsection{Opacities and excitation} 
Recent studies of the Milky
Way and nearby galaxies showed that \ci\, is closely associated with
the CO emission independent of the environment (see e.g. Bennett
\etal\, \cite{bennett94}, Ojha \etal\, \cite{ojha01} for the MW 
and  Gerin \& Phillips \cite{gerin00}, Israel \& Baas \cite{israel02}
for nearby galaxies). Since the critical density
for the \ci\ lines and \aco\, are both $n_{\rm cr} \approx 10^3\,{\rm cm}^{-3}$  
this finding suggests that the transitions arise from the same volume and share
similar excitation temperatures (e.g. Ikeda \etal\, \cite{ikeda02}). 
This allows us to estimate the opacities of the \ci\, transitions via a
LTE analysis assuming optically thick emission in \aco.
Since the \aco\, transition has not been measured in the Cloverleaf we
here adopt ${L}'_{\aco} = 1.1\,{L}'_{\cco}$ (see our discussion on the CO lines). 
Using the values for the line luminosities ($L'$) as listed in Table~\ref{linepara} and
assuming as a start ${T}_{\rm ex}$\,=\,${T}_{\rm dust}$\,=\,50\,K (see below) we find that both
carbon lines are {\it optically thin} ($\tau_{10}$\,=\,0.14, $\tau_{21}$\,=\,0.10).
For a moderate CO opacity (as suggested by B97) optical depth for
the carbon lines are even lower. 
Low optical depths for the \ci\, 
transitions are in line with findings in nearby galaxies
(e.g. Israel \& Baas \cite{israel02}).
As discussed in Stutzki \etal\ (\cite{stutzki97}) the integrated line intensity
ratio $R_{\ci}$ between the \ctwo\, and \cone\, transition can be used in the optically
thin limit to determine directly the \ci\, excitation temperature via 
${T}_{\rm ex}$\,=\,38.8\,${\rm K}/{\rm ln}[2.11/R_{\ci}]$. This
equation is valid if the levels are thermally populated. With $R_{\ci}$\,=\,0.54\,$\pm$\,0.07
(see Table~\ref{linepara}) we find ${T}_{\rm ex}$\,=\,29\,$\pm$\,3\,K. 
For this ${T}_{\rm ex}$ the \ci\, lines are still optically thin 
($\tau_{10}$\,=\,0.15, $\tau_{21}$\,=\,0.11). 
\subsubsection{Neutral carbon and molecular gas mass} The relation between the 
integrated \ctwo\, brightness temperature and the beam averaged \ci\, column density in the
optically thin limit for local galaxies ($z$\,=\,0) is given by
\begin{equation}
N_{\ci} = \frac{8 \pi k \nu_{21}^{2}}{h c^3
A_{21}}\,Q(\tex)\,\frac{1}{5}\,{\rm e}^{T_{2}/
\tex} \int \tmb\,dv
\end{equation}
 where $Q(\tex)=1+3{\rm e}^{-T_{1}/\tex}+5{\rm e}^{-T_{2}/\tex}$ is the \ci\,
 partition function. $T_{1}$\,=\,23.6\,K and $T_{2}$\,=\,62.5\,K are the
 energies above the ground state. 
Correcting for redshift ($\int \tmb\,dv
 \rightarrow (1+z)^{-1}  \int \tmb\,dv$) and accounting for the area of the emission region
 ($\Omega_{\rm s*b}\,D^2_{\rm A}$) yields the total number of carbon atoms in
 the source. Here $\Omega_{\rm s*b}$ denotes the solid angle of the source
 convolved with the telescope beam and $D_{\rm A}$ is the angular size distance to
 the source. Using the definition of the line luminosity
 (e.g. Solomon \etal\, \cite{solomon97}) we can derive the \ci\, mass via
\begin{equation}
\label{eqmci}
M_{\ci} = C\,m_{\ci}\,\frac{8 \pi k \nu_{0}^{2}}{h c^3
A_{21}}\,Q(\tex)\,\frac{1}{5}\,{\rm e}^{T_{2}/
\tex}\,L'_{\ctwo}
\end{equation}
where $m_{\ci}$ is the mass of a single carbon atom and $C$ is the
conversion between pc$^2$ and cm$^2$. Inserting numbers to Eq.\,\ref{eqmci}
yields
\begin{equation}
M_{\ci} = 4.566\times10^{-3}\,Q(\tex)\,\frac{1}{5}\,{\rm e}^{62.5/
\tex}\,L'_{\ctwo} [\msol]
\end{equation}
For $\tex$\,=\,30\,K the mass of neutral carbon thus amounts to
$M_{\ci}$\,=\,2.6\,$\times\,10^7\,m^{-1} \msol$ where $m$ is the magnification by the
gravitational lens.

\noindent Deriving the mass of molecular hydrogen requires an estimate of the 
carbon abundance relative to \hh. We here use M\,82 as a
template to estimate this number. 
Using the results from Stutzki \etal~(\cite{stutzki97}), Walter
\etal~(\cite{walter02}) and Wei\ss~\etal~(\cite{weiss01}) we
obtain [\ci]/[\hh]\,$\approx$\,3\,$\times$\,10$^{-5}$.  
Applying this number to the carbon mass of H1413+117 results in
$M(\hh)$\,$\approx$\,1.4\,$\times$\,10$^{11}\,m^{-1}\,\msol$.
\subsection{Dust emission}
\begin{figure}
   \centering
   \includegraphics[width=7.5cm]{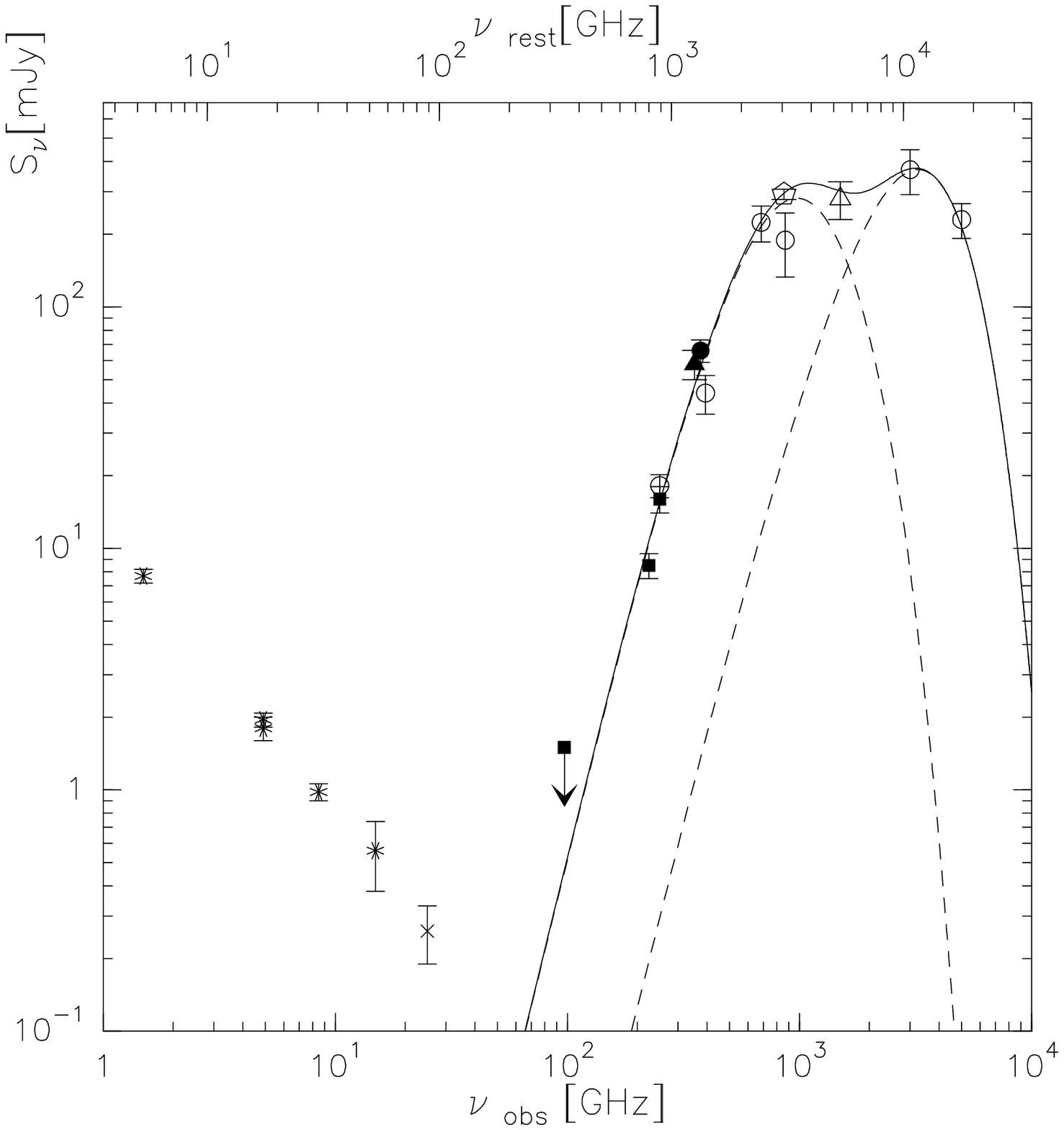}
   \caption{Radio--IR SED of H1413+117. Displayed flux densities were
   taken from Barvainis \& Lonsdale \cite{barlon97} (20, 6, 3, 2\,cm, {\it
   stars}), Solomon \etal\, \cite{solomon03} (1.2\,cm {\it
   cross}), this work (3\,mm, 1.3\,mm, 1.2\,mm, {\it filled squares}),
   Barvainis \etal\, \cite{barvainis95} (1.2\,mm, 760\,\microns,
   438\,\microns, 345\,\microns, 100\,\microns, 60\,\microns, {\it
   circles}), Barvainis \& Ivison \cite{barivi02} (850\,\microns, {\it filled triangle}), 
   Hughes \etal\, \cite{hughes97} (800\,\microns, {\it filled
   circle}), Benford \etal\, \cite{benford99} (350\,\microns, {\it pentagon}) 
   and Rowan-Robinson \cite{rowan00} (200\,\microns, {\it triangle}).   
   The dashed lines show the thermal dust continuum emission for a 50 and 115\,K 
   dust component. The solid line is the total emission from both components.
   }
\label{sed}
\end{figure}

The mm--wave continuum measurements combined with sub-mm, ISO and IRAS data
(for references see caption of Fig.\,\ref{sed}) 
are well fitted with a 2-component dust model. To account for the
steep increase of the observed flux with frequency in the observed mm regime we
used $\kappa_{\rm d}(\nu_{\rm r})$\,=\,10\,$(\nu_{\rm r}/1\,{\rm THz})^\beta\,{\rm
cm^{2}\,g^{-1}}$ with $\beta$\,=\,2 (Downes \& Solomon \cite{downes98}).
For the source solid angle of the lensed dust distribution we
used $\Omega_{\rm S}$\,=\,$\pi\,m\,\theta^2/4$ with
an angular size of $\theta$\,=\,0.17$''$ and a magnification of
$m$\,=\,11 for both components (Venturini \& Solomon \cite{venturini03}).
Masses and dust temperatures for both components were determined by 
fitting those 4 free parameters to the 12 data points above the 97.2\,GHz
upper limit in Fig.\,\ref{sed}. For the cold dust component 
we find $T_{\rm cold}$\,$\approx$\,50\,($\pm$\,2)\,K and 
$M_{\rm cold}$\,$\approx$\,6.7\,$\times$\,10$^8\,m^{-1}\,\msol\,(\pm\,12\%)$. 
The warm component is characterized by
$T_{\rm warm}$\,$\approx$\,115\,($\pm$\,10)\,K and 
$M_{\rm warm}$\,$\approx$\,3.9\,$\times$\,10$^6\,m^{-1}\,\msol\,(\pm\,60\%)$.  
This implies $L_{\rm FIR}\approx2.4\,\times\,10^{13}\,m^{-1}\,\lsol$ 
($L_{\rm FIR}$  as defined by Helou \etal\, \cite{helou85}).
The fit results are displayed in Fig.\,\ref{sed}. 
Parameters derived for the cold gas component are in agreement with
previous studies (e.g. Hughes \etal\, \cite{hughes97}, Benford \etal\, \cite{benford99}).
Using a gas to dust mass ratio of 150 (e.g. Dunne \etal\,
\cite{dunne00}) the total \hh\, mass in the cold
component is $M(\hh)$\,$\approx$\,1.0\,$\times$\,10$^{11}\,m^{-1}\,\msol$ in 
agreement with the mass determined from \ci. Note that a larger angular size
of the cold component leads to lower temperatures for $T_{\rm cold}$,
similarly a smaller size for the warm component will yield higher $T_{\rm warm}$.
\subsection{CO lines}
Our \cco\, flux density is about 30\% higher than the value reported
by B97. 
The flux difference has important implications for the interpretation of the
CO $L'$ ratios: now the 3--2, 4--3, 5--4, 7--6 CO line
luminosities (see footnote in Table\,\ref{linepara}) are decreasing
with increasing $J$. So the data suggest CO is {\it subthermally excited},
at least for ${J}$\,$>$\,3. The minimum kinetic temperature therefore 
can be lower than the lower limit of 60\,K given in B97. We have
reanalyzed the CO ratios using our 3--2 flux and taking the background radiation of
${T}_{\rm CMB}$\,=\,9.6\,K into account. Due to the high 7--6 line luminosity
relative to 5--4 not all CO lines can be fitted with a
single gas component LVG model. If only the 3--2, 4--3 and 5--4 lines
are taken into account LVG solutions can be found for $\tkin$\,$\ge$\,10\,K. 
Since the kinetic temperature is expected to be close to the
excitation temperature of neutral carbon (Israel \& Baas \cite{israel02})
we fixed the $\tkin$ to 30\,K. In a LVG space of [CO]/grad(V) from $1\times10^{-5}$ to
$2\times10^{-4}$ and log($n_\hh$) 1.8 to 5.5, the allowed \hh\,density ranges
from 3.2\,$\le$\,log($n_\hh$)\,$\le$\,4.0 with the higher limit corresponding to
the lowest abundance per velocity gradient and vice versa.
Independently of the abundance per velocity gradient or
the density we find that for \tkin\,=\,30\,K the \aco\, transition will be 
brighter than \cco\, with ${L}'_{\cco}/{L}'_{\aco}$\,$\approx$\,0.9, 
consistent with the lower limit of 0.79 determined by Tsuboi \etal\, (\cite{tsuboi99}). 
Using this line ratio and $M_{\hh}/L'_{\rm CO}$\,=\,0.8\,$\msol\, ({\rm
K}\kms\,{\rm pc}^2)^{-1}$ (Downes \& Solomon \cite{downes98}) we get 
$M(\hh)$\,$\approx$\,1.6\,$\times$\,10$^{11}\,m^{-1}\,\msol$ in agreement
with the result for \ci\, and the dust. Using \tkin\,=\,$T_{\rm dust}$\,=\,50\,K
does not alter the results significantly. The predicted 7--6/3--2
ratio ranges between 0.05\,--\,0.1, much lower than the observed ratio of
0.6. A large fraction of the 7--6 line may therefore arise
from much warmer regions such as the warm dust component. This is also supported by a minimum CO(7--6)
excitation temperature of $\approx$\,75\,K that can be derived from the observed
7--6 brightness temperature. We do not think that selective
magnification would significantly alter this analysis, because
detailed models of the Cloverleaf lens show that the lens
magnification is insensitive to the size of the mm--line emitting
region (Venturini, private communication).
\section{Discussion}  
The mass estimates from \ci, dust, and CO give remarkably similar
results given that each method uses independent assumptions
(carbon abundance, gas to dust mass ratio, CO conversion factor).
Correcting for magnification we find $M(\hh)$\,$\approx$\,1.2\,$\pm$\,0.3\,$\times$\,10$^{10}\,\msol$.

\noindent Compared to the physical conditions of the molecular gas determined in 
other dusty quasars (\tkin\,$\approx$\,50--70\,K, $\nhh$\,$\approx$\,10$^4$,
see e.g. Downes \etal\, \cite{downes99}, 
Carilli \etal\, \cite{carilli02a}) and local ULIRGs
(Downes \& Solomon \cite{downes98}) the values determined above for
H1413+117 argue for a more diffuse and cooler gas component. 
Judging from the excitation temperature of \ci, the excitation of the
molecular gas is between conditions found in M\,82's center ($T_{\rm ex,\ci}$\,= \,50\,K)
and the mean value over the Galactic Center ($T_{\rm
ex,\ci}$\,=\,22\,K) (Stutzki \etal\, \cite{stutzki97}).
The continuum to line ratio of 
$L_{\rm FIR}/L'_{\rm
CO}$\,$\approx$\,120\,\lsol\,(K\,\kms\,pc$^2)^{-1}$ 
is within the range found in local ULIRGs (Solomon \etal\,
\cite{solomon97}) but lower than values estimated for dusty
quasars (e.g. Carilli \etal\, \cite{carilli02a}, Walter \etal\,
\cite{walter03}). All this implies that large amounts of molecular gas are
less affected by the heating from the AGN and/or nuclear star formation than
the region emitting in CO(7--6). A possible explanation for this finding is 
that the molecular disk extends beyond the central region seen in
CO(7--6) to a zone of more moderately excited molecular gas that
dominates the global emission in \ci\, and the low $J$ CO lines.
\begin{acknowledgements}
 We would like to thank R. Neri and R. Zylka for carrying out the observations
 at the PdBI and the 30\,m telescope and S. Venturini for discussions
 on the lens model. IRAM is supported by
 INSU/CNRS (France), MPG (Germany) and IGN (Spain). 
\end{acknowledgements}

\end{document}